\UseRawInputEncoding
\pdfoutput=1

\documentclass[twocolumn]{revtex4}
\newlength{\widthfigcolumn}
\newlength{\widthfigpage}
\usepackage{graphicx}
\usepackage[caption=false, subrefformat=parens,labelformat=parens]{subfig}

\usepackage[]{amsmath}
\usepackage[]{algorithm}
\usepackage[noend]{algpseudocode}
\alglanguage{pseudocode}
\usepackage{algorithmicx,eqparbox,array}
\usepackage{mathtools}
\usepackage{booktabs}
\usepackage{amsthm}

\usepackage{placeins}

\algnewcommand{\LongComment}[1]{\hfill// \begin{minipage}[t]{\eqboxwidth{COMMENT\thealgorithm}}#1\strut\end{minipage}}

\begin{document}
	
\title[Investigation of the Bending Behavior in Silicon Nanowires: A Nanomechanical Modeling Perspective]{Investigation of the Bending Behavior in Silicon Nanowires: A Nanomechanical Modeling Perspective}

\author{Sina Zare Pakzad}
\affiliation{Department of Mechanical Engineering, Ko\c{c} University, 34450 Sariyer, Istanbul, Turkey}
\email[]{szarepakzad18@ku.edu.tr}

\author{Mohammad Nasr Esfahani}
\affiliation{School of Physics, Engineering and Technology, University of York, York, YO10 5DD, UK}
\email[]{mohammad.nasresfahani@york.ac.uk}

\author{B. Erdem Alaca}
\affiliation{Department of Mechanical Engineering \& n${^2}$STAR-Ko\c{c} University Nanofabrication and Nanocharacterization Center \& Ko\c{c} University Surface Technologies Research Center (KUYTAM), Ko\c{c} University, 34450 Sariyer, Istanbul, Turkey}
\email{ealaca@ku.edu.tr}

\begin{abstract}

Nanowires play a pivotal role across a spectrum of disciplines such as nanoelectromechanical systems, nanoelectronics, and energy applications. As nanowires continue to diminish in dimensions, their mechanical characteristics are increasingly influenced by surface attributes. This research delves into the elastic properties of silicon nanowires, which vary in dimensions and crystal orientations, employing diverse nanomechanical models. Through molecular dynamics simulations and the application of five models namely, Heidelberg, Hudson, Zhan, SimpZP, and ExtZP, the analysis of force-deflection responses reveals the intricate interplay among elastic properties, crystal orientation, and the chosen model. To accurately account for unreconstructed states, surface properties are examined for different silicon surface orientations. Each crystal orientation exhibits distinct trends, with surface properties showing a substantial influence on estimated elasticity. A thorough grasp of surface effects and anisotropic properties is crucial for unraveling nanowire mechanics. This study drives both the understanding of size-dependent behaviors and the enhancement of insights into nanoscale surface characteristics of silicon nanowires.

\end{abstract}
	
\maketitle

\section{Introduction}

Nanowires (NWs) hold great promise as fundamental components for advancing future nanoelectromechanical systems \cite{bachtold2022mesoscopic}, nanoelectronics \cite{li2023sub, zhang2021nanowire}, with notable applications in highly sensitive sensors \cite{karimzadehkhouei2023silicon, akbari2020silicon, maita2023revealing} and cutting-edge energy technologies \cite{panda2022piezoelectric}. The remarkable advancements in fabrication methods \cite{wang2016solution, liang2023high}, technological applicability \cite{alaca2023piezoresistive}, and multi-scale modeling techniques \cite{momeni2020multiscale, zare2023nanomechanical} have sparked great interest in the extensive characterization of NWs. The miniaturization of NWs leads to a significant increase in the ratio of surface area to volume giving rise to remarkable physical properties distinguishing them from bulk structures \cite{nasr2019review, wang2017mechanical, huang2023remarkable}. In this context, the surface properties of NWs assume a notably significant role and are regarded as pivotal factors that influence their mechanical attributes \cite{muller2004elastic, zare2023nanomechanical, pakzad2023role}. Furthermore, intrinsic effects such as the stresses generated in NWs during fabrication, assembly, and surface interactions are recognized as dominant parameters that have significant effects on the performance and properties of NWs \cite{zare2023nanomechanical, nasr2022effect, fedorov2019generalized, li2021impact, urena2013raman, yokogawa2017evaluation, lorenzoni2016evaluating, dolabella2020real, spejo2020non}. Nanomechanical modeling techniques are developed as a means to interpret the behavior of NWs at the nanoscale. The classical surface elasticity theory proposed by Gurtin and Murdoch \cite{gurtin1975continuum} has found extensive application in studying the size-dependent mechanical properties of NWs. As dimensions are further scaled down, various surface models are incorporated into nanomechanical modeling approaches in an attempt to elucidate the deformation behavior at the nanoscale regime \cite{he2008surface, song2011continuum, chen2006derivation}. In this regard, different formulations of the Young-Laplace model (YL) have been utilized in numerous studies to interpret the properties of NWs. These include the original formulation of YL \cite{chen2006derivation}, the modified YL \cite{he2008surface, zare2021new}, and the extended YL \cite{song2011continuum, zare2023nanomechanical}. Furthermore, various core-shell models have been developed as structural frameworks that comprise a core with the modulus of the bulk material and a surface shell \cite{yao2012surface,xu2016modified, xiao2022surface}. These models are utilized to interpret and analyze the impact of surface properties on the overall behavior of the NWs \cite{li2022surface, miller2000size, pakzad2023role}.

Silicon (Si) NWs have been extensively investigated as fundamental components in semiconductor manufacturing due to their unique properties \cite{nasr2019review, wang2017mechanical, ye2019last}. Despite numerous studies on the size-dependent mechanical properties of Si NWs, conflicting reports have led to the need for a more detailed examination of the scale effect \cite{zare2023nanomechanical, pakzad2023role, yang2022review}. Specifically, the size-dependent modulus of elasticity of Si NWs has been addressed using various computational and experimental methods, resulting in discrepancies regarding the scale effect \cite{nasr2019review, yang2022review, nasr2019review}. The importance of nanomechanical model selection is highlighted in a recent study, where deviations of up to 85 GPa are attributed to the specific choice of the model \cite{zare2023nanomechanical}. Considering the challenges in fabrication, testing, and characterization, modeling remains a critical step in the successful interpretation of results \cite{zare2021new, zare2023nanomechanical}. The use of bending experiments, aided by advancements in testing methods and \emph{in-situ} approaches, has gained significant importance in revealing the properties of Si NWs \cite{espinosa2013situ}. Within this framework, the development of nanomechanical models that account for large deflections observed in Si NWs has been ongoing \cite{heidelberg2006generalized, ngo2006ultimate, zare2023nanomechanical, zare2021new}. Previous nanomechanical models, such as those proposed by Heidelberg \emph{et al.} \cite{heidelberg2006generalized} for large deflection due to axial extension, extended with initial tension effects by Hudson \emph{et al.} \cite{hudson2013measurement}, incorporating surface effects based on modified YL model by Zhan \emph{et al.} \cite{zhan2012modified}, and combined effects by Zare Pakzad \emph{et al.} \cite{zare2021new}, make this research area particularly intriguing for further exploration. Introducing the ExtZP model \cite{zare2023nanomechanical}, the recent multi-scale approach presents a comprehensive formulation that accounts for large deflections, incorporates anisotropic surface properties using the extended YL model, and considers intrinsic terms encompassing effects arising from surface and initial tension conditions. Moreover, the modeling of the surface properties goes beyond mathematical formulation and also requires an examination of surface state and orientation-dependent properties, such as surface stress and surface elasticity constants. Moreover, recent studies by Zare Pakzad \emph{et al.} \cite{zare2023nanomechanical, pakzad2021molecular} propose a reliable approach to compare unreconstructed and native oxide surface states for (100)- and (110)-Si surfaces. As surface effects continue to gain significance, molecular dynamics (MD) and density functional theory (DFT) simulations are playing vital roles in revealing the size-dependent trends in the elastic properties of Si NWs \cite{pakzad2023role, pakzad2021molecular, lee2007first, zare2023mechanical}. In line with this, MD simulations are used to subject Si NWs to tensile \cite{xu2019molecular, kang2007brittle, pakzad2023role, zare2023mechanical}, vibrational \cite{kim2006molecular, park2005molecular}, and bending tests \cite{zhuo2018atomistic, ilinov2014atomistic}, providing valuable insights into their mechanical properties.

The objective of this study is to explore the elastic properties of Si NWs through the application of nanomechanical modeling, considering different critical dimensions and crystallographic orientations \cite{heidelberg2006generalized, hudson2013measurement, zhan2012modified, zare2021new, zare2023nanomechanical} The models are utilized to interpret the large deflection bending deformations observed in three-point bending tests, allowing for the acquisition and comparison of the modulus of elasticity of Si NWs. The surface properties of Si thin films with different crystal orientations at their free surface are calculated. Furthermore, the calculated intrinsic stresses generated in Si NWs are determined. These results are utilized to comprehensively examine and compare the effectiveness of various nanomechanical models. The study establishes a multi-scale framework for developing a strategy to investigate the size-dependency of Si NW properties, particularly considering unreconstructed surface states. Through this work, valuable insights into the surface characteristics of Si NWs are provided, contributing to an improved understanding of their mechanical properties at the nanoscale.

\section{Materials and Methods}
This section encompasses four segments, which focus on the methods used to interpret the selection of nanomechanical models and their effects on the calculated elastic responses of Si NWs. These include the analytical constitution of the models, formulations for effective bending rigidity, three-point bending simulations of Si NWs using MD, and atomistic simulations of surface properties. The methods are utilized to interpret the elastic response of Si NWs with varying sizes and crystal orientations through different nanomechanical models. The order in which these methods are described is as follows: Section \ref{sec:2.1} provides details about the existing nanomechanical models employed to interpret the bending response in NWs. Section \ref{sec:2.2} discusses the integration of effective modulus of elasticity definitions into the models, along with their respective formulations that incorporate surface analogy, elucidating the bending rigidity details. Section \ref{sec:2.3} delves into the bending simulations performed on Si NWs via MD method. Finally, Section \ref{sec:2.4} presents the atomistic method employed to calculate surface properties (surface stress and surface elasticity constants) for different orientations of unreconstructed Si film.

\subsection{Nanomechanical Models}
\label{sec:2.1}

\begin{figure*}[ht]
	\centering
	\includegraphics[width=0.9\textwidth]{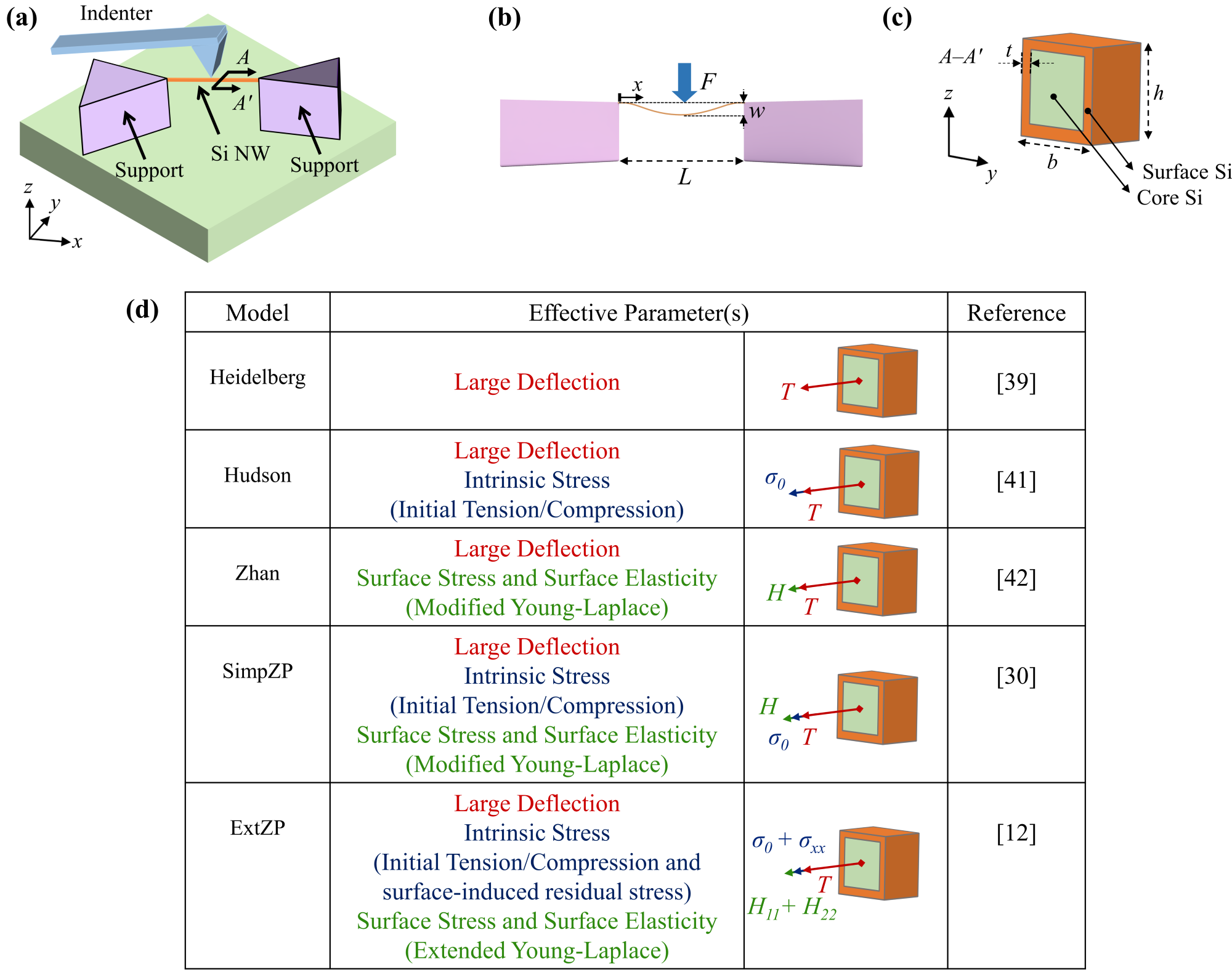}
	\caption{\label{fig:fig1} (a) Schematic representation of Si NW, clamped by support regions and subjected to the force applied by an indenter. (b) Schematic of a NW of length $L$ under bending test with applied load $F$, and subsequent deformation $w$ at its center. (c) Cross-sectional representation of the NW with associated height ($h$), width ($b$), and surface layer thickness ($t$). (d) Effective terms of different nanomechanical models are given here.}
\end{figure*}

The section begins by introducing the Heidelberg model, which represents the most general version of large deflection models utilized in the interpretation of bending deformations in NWs \cite{heidelberg2006generalized}. Following that, Section \ref{sec:2.1.2} introduces an extended version of the Heidelberg model called the Hudson model \cite{hudson2013measurement}, which incorporates intrinsic stress induced by initial tension/compression. In Section \ref{sec:2.1.3}, Zhan \emph{et al.}'s modification of the Heidelberg model \cite{zhan2012modified} is discussed, including the modified YL surface model to incorporate surface properties. Section \ref{sec:2.1.4} explains the large deflection model proposed by Zare Pakzad \emph{et al.} \cite{zare2021new}, referred to as SimpZP. This model incorporates both intrinsic stress induced by initial tension/compression and the modified YL surface model. Finally, in Section \ref{sec:2.1.5}, a recently developed nanomechanical model, known as the ExtZP model \cite{zare2023nanomechanical}, is presented. It integrates the extended YL model and intrinsic stresses induced by both surface effects and initial tension/compression. These nanomechanical models enable the prediction of the modulus of elasticity ($E$) for Si NWs by employing different sets of parameters as inputs. Figure~\ref{fig:fig1} (a) provides an overview, schematically illustrating the effective parameters associated with each model, including large deflection, intrinsic stress, and surface properties.

\subsubsection{Heidelberg Model}
\label{sec:2.1.1}

The Heidelberg model presents the most general method for interpreting the bending behavior of Si NWs when subjected to large deformation during a three-point bending test. The governing equation for a NW of length $L$ and of rectangular cross section loaded at its midsection by a point load $F$ (Figure~\ref{fig:fig1} (a)) is given in Equation \ref{eqn1}.

\begin{equation}
	\label{eqn1}
	EI\frac{\partial^{4} w}{\partial x^{4}} - T\frac{\partial^{2} w}{\partial x^{2}}  = \frac{F}{2}\big(-\delta(x) + 2\delta(x-\frac{L}{2})-\delta(x-L)\big)
\end{equation}

In Equation~\ref{eqn1}, $w$ represents the transversal displacement in the $z$-direction, $x$ denotes the longitudinal axis of the beam, $I$ represents the moment of inertia, $T$ signifies the overall tension along the NW, and $\delta$ is the Dirac delta function, as depicted in Fig.~\ref{fig:fig1} (b). The first term in Equation~\ref{eqn1} arises from the bending of the beam, while the second term is associated with the stretching along the axial direction of the beam, often referred to as large deflection in nanomechanics. The axial tension term can be expressed as Equation~\ref{eqn2}.

\begin{equation}
	\label{eqn2}
	T =  \frac{EA}{2L} \int_{0}^{L} (\frac{\partial w}{\partial x})^2 dx
\end{equation}

By integrating Equation~\ref{eqn1} and Equation~\ref{eqn2} with respect to $x$, and taking into account the usual clamped boundary conditions (B.C.s) at both ends of the NW, as well as the cross-sectional area denoted as $A$, Equation~\ref{eqn3} is obtained.

\begin{equation}
	\label{eqn3}
	F = \frac{192EI}{L^3} w(\frac{L}{2})f(\alpha)
\end{equation}

Equation~\ref{eqn3} is commonly known as the linear beam deflection approximation, with the relationship $f(\alpha)=1$. However, when a precise solution method is employed, the function $f(\alpha)$ yields Equation~\ref{eqn4}.

\begin{equation}
	\label{eqn4}
	f(\alpha) = \frac{\alpha}{48-\frac{192\tanh(\sqrt{\alpha}/4)}{\sqrt{\alpha}}}
\end{equation}

The value of $\alpha$, which is connected to the maximum deflection through Equation~\ref{eqn5} is utilized to calculate $f(\alpha)$ in order to predict the modulus of elasticity of Si NWs using Heidelberg model. 

\begin{align}
	\label{eqn5}
	\begin{split}
		\big(\frac{\alpha\cosh^2(\frac{\sqrt{\alpha}}{4})}{2 +  \cosh(\frac{\sqrt{\alpha}}{2}) - 6 \frac{\sinh(\frac{\sqrt{\alpha}}{2})}{\sqrt{\alpha}}}\big) &\big(1-4\frac{\tanh(\sqrt{\alpha}/4)}{\sqrt{\alpha}}\big)^2 \\
		&= \Delta Z_{center}^2 \big(\frac{E}{I}\big)
	\end{split}	
\end{align}

\subsubsection{Hudson Model}
\label{sec:2.1.2}

According to the Hudson model \cite{hudson2013measurement}, in order to analytically include residual stresses resulting from initial tension/compression into the Heidelberg model, it is essential to incorporate the term $T_0$, which represents the initial axial force along the NW. This term is added to Equation~\ref{eqn2} with an updated formulation provided in Equation~\ref{eqn6}. The Hudson model utilizes a governing equation that is similar to the Heidelberg model, thereby allowing it to account for the effects of embedded initial stresses.

\begin{equation}
	\label{eqn6}
	T = T_0 + \frac{EA}{2L} \int_{0}^{L} (\frac{\partial w}{\partial x})^2 dx
\end{equation}

The presence of the term $T_0$ in the formulation signifies the residual term that is embedded within the NW, arising from fabrication or assembly-induced stresses. By solving Equation~\ref{eqn1} and Equation~\ref{eqn6} according to the Hudson model, the exact analytical solution provided in Equation~\ref{eqn7} is derived.

\begin{align}
	\begin{split}
\label{eqn7}
	(1-\frac{L^2T_0}{\alpha EI})&(\frac{EA}{T_0 + EA}) \big(\frac{\alpha\cosh^2(\frac{\sqrt{\alpha}}{4})}{2 + \cosh(\frac{\sqrt{\alpha}}{2}) - 6 \frac{\sinh(\frac{\sqrt{\alpha}}{2})}{\sqrt{\alpha}}}\big)\\ &\big(1-4\frac{\tanh(\sqrt{\alpha}/4)}{\sqrt{\alpha}}\big)^2  = \Delta Z_{center}^2 \big(\frac{E}{I}\big)	
	\end{split}
\end{align}

It is worth mentioning that an approximate version of Hudson's model is proposed by Yaish \emph{et al.} \cite{yaish2015three}, where the function $f(\alpha)$ is approximated as Equation~\ref{eqn8}.

\begin{equation}
	\label{eqn8}
	f(\alpha) = 1 + 2.412\times10^{-2} \alpha - 1.407\times10^{-6} \alpha^2
\end{equation}

\subsubsection{Zhan Model}
\label{sec:2.1.3}

The Zhan model \cite{zhan2012modified}, which incorporates a surface model based on the modified YL equation, introduces updates to the formulation of the Heidelberg model. These updates are reflected as surface stress and surface elasticity constant given in Equation~\ref{eqn9}. As the analytical bending formulation incorporates the surface model, the concept of the effective modulus of elasticity, represented as $E^*$, becomes necessary. Consequently, all formulations utilizing surface embedded models consider both the effective modulus of elasticity and the effective moment of inertia ($I^*$). Subsequently, this article will delve into the specifics of the bending rigidity formulation for determining the modulus of elasticity of Si NWs Section \ref{sec:2.2}. Equation~\ref{eqn9} shows the governing equation provided by Zhan model.

\begin{align}
	\begin{split}
		\label{eqn9}
		&(EI)^*\frac{\partial^{4} w}{\partial x^{4}} - T\frac{\partial^{2} w}{\partial x^{2}} - H\frac{\partial^{2} w}{\partial x^{2}} = \\
		& \frac{F}{2}\big(-\delta(x) + 2\delta(x-\frac{L}{2})-\delta(x-L)\big)
	\end{split}
\end{align}

In Equation~\ref{eqn9}, $H$ represents the surface parameter that incorporates the surface stress, $f_{11}$, and the surface elasticity constant, $d_{11}$ with detailed form given in Equation~\ref{eqn10}.

\begin{equation}
	\label{eqn10}
	H = 2hd_{11} \epsilon_{x} +2f_{11}h
\end{equation}

In Equation~\ref{eqn10}, $h$ represents the thickness of the NW, and $\epsilon_{x}$ denotes the axial strain. By solving Equation~\ref{eqn9}, following a similar approach as the previous models, the solution given in Equation~\ref{eqn11} is obtained.

\begin{align}
	\begin{split}
	\label{eqn11}
	&(1-\frac{2f_{11}hL^2}{\alpha(EI)^*}) \big(\frac{\alpha\cosh^2(\frac{\sqrt{\alpha}}{4})}{2 + \cosh(\frac{\sqrt{\alpha}}{2}) - 6 \frac{\sinh(\frac{\sqrt{\alpha}}{2})}{\sqrt{\alpha}}}\big)\\
	&\big(1-4\frac{\tanh(\sqrt{\alpha}/4)}{\sqrt{\alpha}}\big)^2
	= \Delta Z_{center}^2 \big(\frac{(EA)^* + 2d_{11}h}{(EI)^*}\big)
	\end{split}
\end{align}

In this regard, $E^*$ for the Si NW is derived by numerically solving Equation~\ref{eqn11}. The bending rigidity associated with the modified YL model is utilized to predict the modulus of elasticity of the Si NW. Section \ref{sec:2.2} will elaborate on the precise formulation concerning the bending rigidity aligned with Zhan model.

\subsubsection{SimpZP Model}
\label{sec:2.1.4}
The SimpZP model \cite{zare2021new} incorporates both fabrication induced residual stress, $\sigma_{0}$, and surface effects, $H$, via modified YL formulation. In this context, this model leverages the governing terms outlined in Equation~\ref{eqn12}, wherein Equation~\ref{eqn6} replaces Equation~\ref{eqn2} to represent the axial extension term, considering the embedded initial tension/compression component. Essentially, this approach integrates the formulations provided by the Heidelberg, Hudson, and Zhan models, encompassing all the previously discussed terms associated with these models. The solution provided by SimpZP model is given in Equation~\ref{eqn12}.

\begin{align}
	\begin{split}
	\label{eqn12}
		&(1-\frac{L^2T_0}{\alpha(EI)^*})(\frac{(EA)^*}{T_0 + (EA)^*})(1-\frac{2f_{11}hL^2}{\alpha(EI)^*}) \\
&					\big(\frac{\alpha\cosh^2(\frac{\sqrt{\alpha}}{4})}{2 + \cosh(\frac{\sqrt{\alpha}}{2}) - 6 \frac{\sinh(\frac{\sqrt{\alpha}}{2})}{\sqrt{\alpha}}}\big)\big (1-4\frac{\tanh(\sqrt{\alpha}/4)}{\sqrt{\alpha}}\big)^2 \\
& =\Delta Z_{center}^2 \big(\frac{(EA)^* + 2d_{11}h}{(EI)^*}\big)
	\end{split}
\end{align}

In accordance with the SimpZP formulation, the bending rigidity formulation derived from the modified YL model is integrated with this model to provide the relation between $E$ and $E^*$. Similar to Zhan model, in Section \ref{sec:2.2}, the specific formulation regarding the bending rigidity, in alignment with the SimpZP model, will be elaborated upon.

\subsubsection{ExtZP Model}
\label{sec:2.1.5}
The model proposed by Zare Pakzad \emph{et al.} \cite {zare2023nanomechanical}, ExtZP model, represents a multiscale extension of prior models, where the treatment of the surface is enhanced by incorporating two new sets of effective parameters: i) anisotropic surface parameters, $H_{11}$ and $H_{22}$, and ii) surface-induced residual stresses, $\sigma_{xx}$. Consequently, the ExtZP approach represents the most comprehensive method for interpreting the bending response of Si NWs. In this approach, Equation~\ref{eqn13} incorporates the surface terms derived from the extended YL surface model, which are combined with the governing equation presented in Equation~\ref{eqn9} and the initial extension term expressed in Equation~\ref{eqn6}.In this formulation, $b$ represents the NW width.

\begin{align}
	\begin{split}
	\label{eqn13}
	&H = H_{11} + H_{22} + bh\sigma_{xx} = \\
	& 2bf_{11} + 2hf_{22} + 2bd_{11} + 2hd_{22} + bh\sigma_{xx}
	\end{split}
\end{align}

With terms $H_{11}$ and $H_{22}$ defined as effective surface terms embedding the surface constants and geometrical representation of involved side surfaces \cite{song2011continuum} the complete form of ExtZP model can be rewritten as Equation~\ref{eqn14} where $(EI)^*$ represents the effective flexural rigidity. The indices $11$ and $22$ denote the corresponding directions for the involved planar surfaces.

\begin{equation}
	\label{eqn14}
	\begin{split}
		&(EI)^*\frac{\partial^{4} w}{\partial x^{4}} - T\frac{\partial^{2} w}{\partial x^{2}} - 2bf_{11}\frac{\partial^{2} w}{\partial x^{2}} - 2hf_{22}\frac{\partial^{2} w}{\partial x^{2}} \\
		&-2bd_{11}\frac{\partial^{2} w}{\partial x^{2}} - 2hd_{22}\frac{\partial^{2} w}{\partial x^{2}}  -bh\sigma_{xx}\frac{\partial^{2} w}{\partial x^{2}} = \\ 
		&\frac{F}{2}\big(-\delta(x) + 2\delta(x-\frac{L}{2})-\delta(x-L)\big)
	\end{split}
\end{equation}

Expanding upon the terms present in Equation~\ref{eqn14}, there are five terms that account for surface effects. These terms consist of the surface stress components ($f_{11}$ and $f_{22}$), surface elasticity constants ($d_{11}$ and $d_{22}$), and the residual stress induced by the surface ($\sigma_{xx}$). The exact solution for the nanomechanical model is given in Equation~\ref{eqn15}. Section \ref{sec:2.2} will provide the bending rigidity formulation for the ExtZP model, which is based on the extended YL approach.

\begin{equation}
	\label{eqn15}
	\begin{split}
		&(1-\frac{L^2T_0}{\alpha(EI)^*})(\frac{(EA)^*}{T_0 + (EA)^*})(1-\frac{L^2H_{11}}{\alpha(EI)^*}) (1-\frac{L^2H_{22}}{\alpha(EI)^*})\\
		&\big(\frac{\alpha\cosh^2(\frac{\sqrt{\alpha}}{4})}{2 + \cosh(\frac{\sqrt{\alpha}}{2}) - 6 \frac{\sinh(\frac{\sqrt{\alpha}}{2})}{\sqrt{\alpha}}}\big)\big(1-4\frac{\tanh(\sqrt{\alpha}/4)}{\sqrt{\alpha}}\big)^2 = \\
		& \Delta Z_{center}^2 \big(\frac{(EA)^* + 2d_{11}b + 2d_{22}h + bhE^*}{(EI)^*}\big)
	\end{split}
\end{equation}

\subsection{Bending Rigidity Calculation}
\label{sec:2.2}

To estimate the value of $E$, the connection between classical bending rigidity and effective bending rigidity is employed, which relies on surface properties and geometric definitions. The formulation for the effective bending rigidity, as given by Equation~\ref{eqn16}, utilizes a modified YL method \cite{he2008surface, wang2007effects} that aligns with the Zhan and SimpZP models.

\begin{equation}
	\label{eqn16}
	(EI)^* = EI + (E_s)(\frac{bh^2}{2}+\frac{h^3}{6})
\end{equation}

In this context, the ExtZP model requires the calculation of the modulus of elasticity, which relies on the effective bending rigidity formulated using the extended YL model \cite{song2011continuum, zare2023nanomechanical}, as expressed in Equation~\ref{eqn17}.

\begin{equation}
	\label{eqn17}
	(EI)^* = EI + (E_s + \tau_0)(\frac{bh^2}{2}+\frac{h^3}{6})
\end{equation}

These formulations designate the average values of surface stresses ($f_{11}$ and $f_{22}$) and surface elasticity constants ($d_{11}$ and $d_{22}$) as $\tau_0$ and $E_s$, respectively. In the field of nanomechanical models, these terms are commonly known as the surface modulus and initial surface stress, as discussed in the literature \cite{he2008surface}. More detailed information regarding the formulation of previous models \cite{heidelberg2006generalized, zhan2012modified, hudson2013measurement, zare2021new, zare2023nanomechanical} and related expressions for bending rigidity \cite{song2011continuum, he2008surface, wang2007effects} can be found in other sources.

\subsection{Bending Modeling}
\label{sec:2.3}

\begin{figure}[ht]
	\centering
	\includegraphics[width=0.5\textwidth]{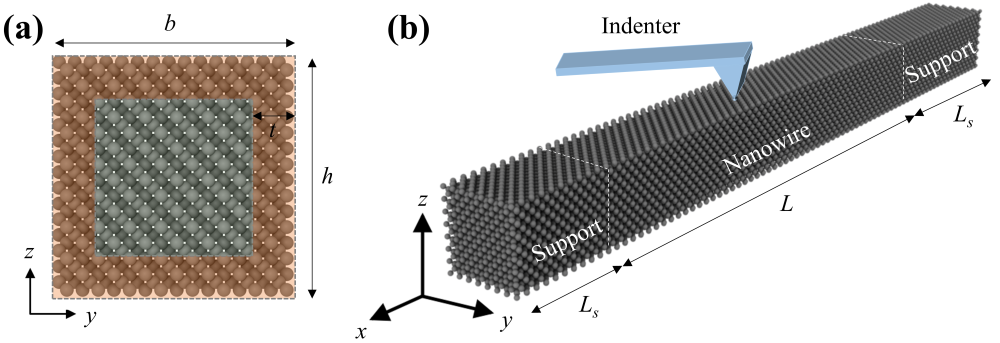}
	\caption{\label{fig:fig2} (a) The initial configuration of Si NW of width, $b$, height , $h$, and surface layer thickness, $t$. (b) The perspective view of a double-clamped NW showing the atomic configuration of a Si NW under 3-point bending simulation using a spherical indenter with associated support regions.The NW length ($L$) and support region length ($L_s$) are shown here.}
\end{figure}

Simulations of three-point bending are conducted on double-clamped Si NWs using MD method \cite{plimpton1995fast}. Figure~\ref{fig:fig2} illustrates the initial atomic configurations of Si NWs, where the width ($h$) and height ($b$) are varied between 2 nm and 5 nm. The Si NWs have a cubic cross-section, with the width being equal to the height. In this regard, the length ($L$) of the Si NW remains constant at 600 nm for different critical dimensions. Furthermore, it should be noted that support regions exist on each side of the Si NW and have a length ($L_s$) of 60 nm. It is crucial to mention that both the length of the NW and the length of the support regions are kept constant across all critical dimensions of the Si NWs. This ensures a length-to-width aspect ratio (AR) greater than 10 for all NW critical dimensions, effectively excluding any potential length-related effects \cite{pakzad2023role}. A constant surface layer thickness ($t$) of 0.5 nm is assumed for all Si NWs studied here. To maintain consistency in cross-section designs and accurate examination of surface constants, it is common to observe rectangular or square geometries in top-down fabricated Si NWs with $<100>$ and $<110>$ orientations \cite{nasr2019review}. On the other hand, circular cross-sections are frequently encountered in bottom-up fabricated Si NWs with $<111>$ and $<112>$ orientations \cite{nasr2019review}. In order to maintain consistency in defining the height and thickness of the Si NWs and to accurately analyze surface constants, square cross-sections have been chosen for all four orientations investigated in this study. This decision ensures that the cross-section designs are uniform across the orientations ($<100>$, $<110>$, $<111>$ and $<112>$), enabling precise comparisons and comprehensive analysis of the Si NWs. To simulate the clamped B.C., the boundary regions at both ends are fixed in all three dimensions ($x$, $y$, and $z$). The non-periodic shrink-wrapped B.C.s are applied in all directions. The Si NWs undergo a quasi-static relaxation process using the conjugate gradient method to reach a local minimum configuration. Subsequently, a relaxation process is performed on all NWs at a temperature of 1.0 K for 40 ps, utilizing a time step of 1 fs. This is accomplished by employing a canonical NVT ensemble with a Nose-Hoover thermostat. The Tersoff-T3 potential \cite{tersoff1988new} as an empirical function composed of two-body terms depending on the local environment is used for the modeling of Si NWs in this work. The associated Si NW stress state, $\pi_{ij}$, is calculated using the virial theorem \cite{zimmerman2004calculation}, given in Equation~\ref{eqn18}.

\begin{equation}
	\label{eqn18}
	\pi_{ij} = \frac{1}{2\Omega_{0}}\bigg[\sum_{\alpha=1}^{N}\sum_{\beta \neq \alpha}^{N} \frac{1}{r^{\alpha\beta}} \frac{\partial{V}(r^{\alpha\beta})}{\partial r} (v_{i}^{\alpha\beta}v_{j}^{\alpha\beta} )\bigg]
\end{equation}

In Equation~\ref{eqn18}, $\Omega_0$ stands for the atomic volume in an undeformed system with $N$ as the total number of atoms where associated atomic volume for Si is calculated via relaxation of related structure with further details given in Ref. \cite{pakzad2021molecular}. Atomic distances between atoms $\alpha$ and $\beta$ are represented as $r_{\alpha\beta}$. $v_{j}^{\alpha}$ stands for the position of atom $\alpha$ along $j$ direction, i.e., $v_{j}^{\alpha\beta} = v_{j}^{\alpha}-v_{j}^{\beta}$ and $V$ represents the interatomic potential. The $F/w$ ratio, which is the ratio of force to deflection, can be determined by analyzing the slope of the initial linear section of the force-deflection response. This ratio is often used to estimate the modulus of elasticity through a linear approach. However, it should be noted that in the case of bending response curves for Si NWs, the displacement of the NW upon loading is expected to be larger than half of the cross-sectional size ($h/2$). As a result, the response is likely to exhibit a non-linear behavior, as observed in prior experimental and MD studies on the bending of NWs \cite{zare2023nanomechanical, pakzad2021molecular}. In this context, the nanomechanical models described in Section \ref{sec:2.1} will be utilized to estimate the modulus of elasticity for Si NWs with various crystallographic orientations and critical dimensions. 

\subsection{Surface Properties Modeling}
\label{sec:2.4}

\begin{figure}[ht]
	\centering
	\includegraphics[width=0.5\textwidth]{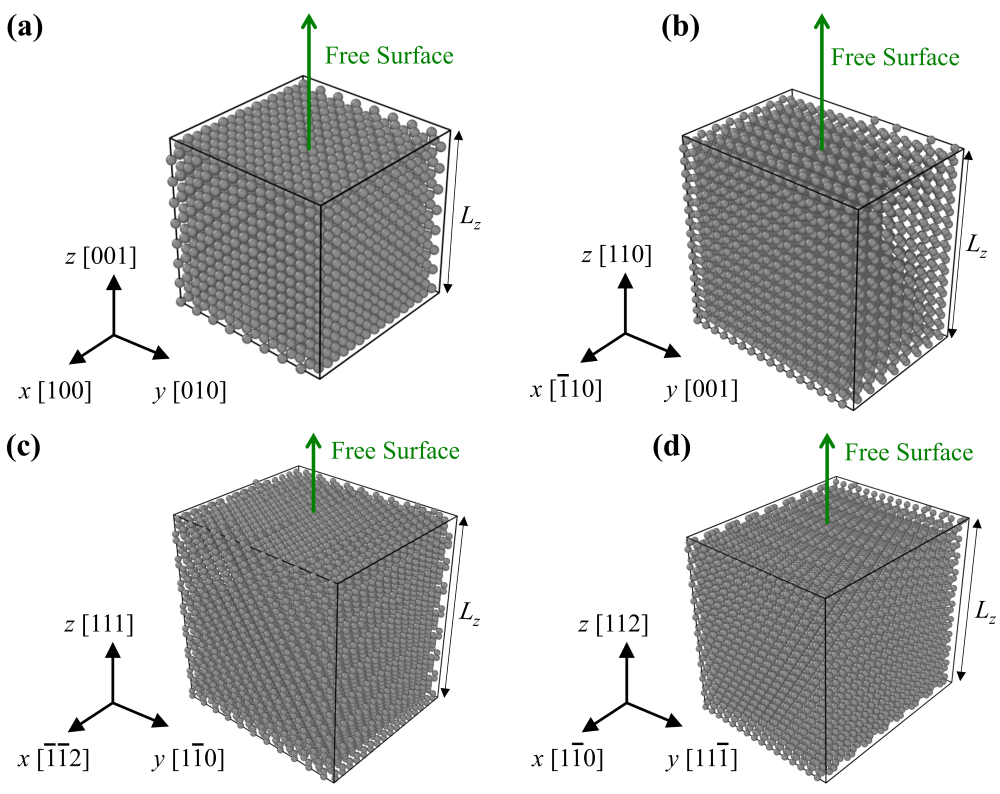}
	\caption{\label{fig:fig3} The atomic structures used for calculating surface properties of unreconstructed Si thin films with different crystallographic orientations and associated free surfaces in the $z$-direction. The depicted structures correspond to Si orientations of (a) Si (100), (b) Si (110), (c) Si (111), and (d) Si (112).}
\end{figure}

To quantify the surface stress components ($f_{11}$ and $f_{22}$) and surface elastic constants ($d_{11}$ and $d_{22}$) for four orientations of Si NWs, Martin's method can be applied to the thin film structures given in Figure~\ref{fig:fig3}. By employing periodic boundary conditions in the $x$- and $y$-directions, an infinite film model with a finite thickness ($L_z$) in the $z$-direction is obtained. The evaluation area ($A_{film}$) consists of two free surfaces facing each other, and the film must be sufficiently thick to prevent interaction between these two surfaces. The dimensions, crystallographic orientations, numbers of atoms, and evaluation areas are listed in Table~\ref{tab:table1}. The atomic structures are designed to be 8 times the lattice constants of Si along each crystallographic direction. This ensures an appropriate thickness for the film, allowing for accurate calculations of surface stresses and surface elastic constants.

\begin{table*}\centering
	\caption{\label{tab:table1}Dimensions, number of atoms and evaluation areas of unreconstructed Si thin film structures with associated crystallographic orientations.}
	\begin{tabular}{ccccccccc}
		\toprule
		Structure& \multicolumn{3}{c}{Si Orientation} & \multicolumn{3}{c}{Dimensions [nm]}  & No. of atoms & $A_{film}$ [nm$^2$]\\
		\hline\addlinespace[2pt]
		\hline\addlinespace[2pt]
		& $x$ & $y$ & $z$	& $x$ & $y$ & $z$ & & \\
		\cmidrule(l){2-4}
		\cmidrule(l){5-7}
		\cmidrule(l){8-9}
		\hline\addlinespace[2pt]
		Si $(100)$ 		 & {[100]}  & [010] & [001] & 			4.35 & 4.35 & 4.35 & 4224  & 37.8  \\
		\hline\addlinespace[2pt]
		Si $(110)$     	& [$\bar{1}$10]  & [001] & [110] & 	6.15 & 4.35 & 6.15 & 8448 & 53.5  \\
		\hline\addlinespace[2pt]
		Si $(111)$ 		 & [$\bar{1}$$\bar{1}$2] & [1$\bar{1}$0] & [111] & 	7.10 & 6.15 & 7.52 & 16378 & 
		 87.3 \\
		 \hline\addlinespace[2pt]
		Si $(112)$  	 & [1$\bar{1}$0]  & [11$\bar{1}$] & [112] & 	6.15 & 7.52 & 7.10 & 16356  & 106.8  \\
		\hline\addlinespace[2pt]
		\bottomrule		
	\end{tabular}
\end{table*}

To determine the surface stress components, the structures are initially equilibrated at a temperature of 0.01 K using NPT dynamics. Afterward, the total energy is minimized using the steepest descent and conjugate gradient techniques. The surface stress components are then calculated by assuming a film thickness of $L_z$, and are formulated according to Equation~\ref{eqn19}. In this equation, the parameter $\eta$ represents the average strain resulting from the relaxation process \cite{izumi2004method, pakzad2021molecular, zare2023nanomechanical}. The Tersoff-T3 potential \cite{tersoff1988new} is used for the modeling of Si thin films.

\begin{equation}
	\label{eqn19}
	f_{ij}(\eta) = f_{ij}^{film}(\eta) - f_{ij}^{bulk}(\eta)
\end{equation} 

The estimation of bulk surface stress requires a separate analysis using the bulk model. The surface stress for the film model can be determined by utilizing the stress tensor definition, denoted as $\sigma_{ij}^{film}$, as described in Equation~\ref{eqn20}. For more comprehensive information regarding the stress tensor of thin films, additional details can be found elsewhere \cite{izumi2004method, pakzad2021molecular, zare2023nanomechanical}.

\begin{equation}
	\label{eqn20}
	f_{ij}^{film}(\eta) = \sigma_{ij}^{film}(\eta) \frac{L_z}{2}
\end{equation} 

The surface elastic constants ($d_{ijkl}$) are defined as the second-order derivatives of surface energy with respect to strain under plane-stress conditions. To calculate the elastic constants for both the bulk ($C_{ijkl}^{bulk}$) and film ($C_{ijkl}^{film}$) models, specific considerations are made. For the bulk model, the elastic constants are defined under plane-stress conditions ($C_{ijkl}^{bulk, plane}$), which can be determined using the relationships provided in Equation~\ref{eqn21}, Equation~\ref{eqn22}, and Equation~\ref{eqn23} \cite{izumi2004method, pakzad2021molecular, zare2023nanomechanical}.

\begin{equation}
	\label{eqn21}
	C_{11}^{bulk, plane} = C_{11} - \frac{C_{13}^2}{C_{33}}
\end{equation}

\begin{equation}
	\label{eqn22}
	C_{22}^{bulk, plane} = C_{22} - \frac{C_{12}^2}{C_{33}}
\end{equation}

\begin{equation}
	\label{eqn23}
	C_{12}^{bulk, plane} = C_{12} - \frac{C_{12}C_{13}}{C_{33}}
\end{equation}

The relationship between the plane-stress elastic constants of the bulk and film models and the surface elastic constants is expressed by Equation~\ref{eqn24} \cite{izumi2004method, pakzad2021molecular, zare2023nanomechanical}. To obtain elastic constants, infinitesimal strains are applied to the simulation box in different directions and the resulting variation of the energy is measured.

\begin{equation}
	\label{eqn24}
	C_{ijkl}^{film, plane} = C_{ijkl}^{bulk, plane} - \frac{2}{L_z} d_{ijkl}
\end{equation}

\section{Results}
\label{sec:results}

The section begins by establishing surface parameters, such as surface stresses and surface elastic constants, for different orientations of unreconstructed Si thin films (Section \ref{sec:3.1}). Next, the bending simulation results for Si NWs modeled using MD with varying critical dimension and crystallographic orientations are presented in Section \ref{sec:3.2}. Subsequently, the force-deflection relationships resulting from the simulations are examined to extract the modulus of elasticity of the Si NWs. The modulus of elasticity results are interpreted using different models from the existing literature (Section \ref{sec:3.3}).

\subsection{Surface Properties of Si}
\label{sec:3.1}

Despite numerous studies employing atomistic \cite{pakzad2021molecular, pishkenari2017characterization, izumi2004method, xu2019molecular, park2008surface}, DFT \cite{melis2015lattice, melis2016density, lee2007first, jung2017effects}, finite element \cite{quagliotti2013finite, delph2008near, payne1989total, miyamoto1994comparative, alerhand1988spontaneous}, and experimental \cite{pennelli2012correlation} approaches to investigate the surface constants of Si, a consensus regarding surface stress profiles and the subsequent calculation of constants for different crystalline orientations and surface conditions is lacking. A recent study has provided insights into the surface constants of unreconstructed and oxidized Si thin films offering valuable analysis regarding the surface properties of crystalline Si with different surface states \cite{zare2023nanomechanical}. In line with that investigation, the unreconstructed Si thin films with (100), (110), (111), and (112) free surface orientations shown in Fig.~\ref{fig:fig3} are examined for the calculation of surface properties including surface stresses, $f_{11}$ and $f_{22}$, and surface elasticity constants, $d_{11}$ and $d_{22}$. The methodology described in Section~\ref{sec:2.4} is employed to analyze nanoscale Si thin films. The validity of this approach has been previously verified on various Si surfaces, including amorphous and unreconstructed structures, using different interatomic potentials. These previous studies, conducted through MD and DFT, have demonstrated good agreement with the results obtained from the current method \cite{shchipalov2000surface, miller2000size}. To calculate surface stress, Equation~\ref{eqn19} is utilized, taking into account the film thickness ($L_z$), and the resulting planar surface stresses are presented in Table~\ref{tab:table2}. Additionally, the surface elasticity constants are computed using Equations~\ref{eqn21}, \ref{eqn22}, and \ref{eqn23}, with the values also provided in Table~\ref{tab:table2}. The local stresses along the thin film structures are calculated using Equation~\ref{eqn18}, considering the estimated atomic volume. The surface elastic constants are obtained for the film structures using limited thicknesses, while the bulk models are separately examined. Equation~\ref{eqn24}, with the associated film thickness of $L_z$, is employed for this purpose. Table~\ref{tab:table2} further illustrates the average surface stress and surface elasticity constants for each orientation of the free surface. Remarkably, the obtained surface stress and surface elasticity constants for Si (100) and Si (110) crystal orientations align precisely with prior reports \cite{zare2023nanomechanical, pishkenari2017characterization, izumi2004method, pakzad2021molecular, shchipalov2000surface}. To the best of our knowledge, there is currently no existing study that provides surface constants for (111) and (112) Si thin films. The surface properties obtained here will be utilized in Section~\ref{sec:3.2} to interpret the bending response of Si NWs using different surface formulations integrated into nanomechanical models.

\begin{table*}\centering
	\caption{\label{tab:table2} Surface stress and surface elastic constants calculated for unreconstructed Si structures with associated crystallographic orientations.}
	\begin{tabular}{cccccccccc}
		
		\toprule
		Structure& \multicolumn{3}{c}{Si Orientation} & \multicolumn{3}{c}{Surface Stress [N/m]}  & \multicolumn{3}{c}{Surface Elasticity [N/m]}\\
		\midrule
		\hline\addlinespace[2pt]
		\hline\addlinespace[2pt]
		& $x$ & $y$ & $z$	& $f_{11}$ & $f_{22}$ & $f_{avg}$	& $d_{11}$ & $d_{22}$ & $d_{avg}$\\
		\cmidrule(l){2-4}
		\cmidrule(l){5-7}
		\cmidrule(l){8-10}
		\hline\addlinespace[2pt]
		Si $(100)$ 		 & {[100]}  & [010] & [001] & 		-0.88 & -0.88 & -0.88 & -8.77 & -8.77 & -8.77  \\
		\hline\addlinespace[2pt]
		Si $(110)$     	& [$\bar{1}$10]  & [001] & [110] & 	-1.57 & -0.95 & -1.26 & +3.40 & -15.10 & -5.85 \\
		\hline\addlinespace[2pt]
		Si $(111)$ 		 & [$\bar{1}$$\bar{1}$2] & [1$\bar{1}$0] & [111] & 	-0.88 & -1.23 & -1.06 & -18.67 &  +8.42 & -5.12\\
		\hline\addlinespace[2pt]
		Si $(112)$  	 & [1$\bar{1}$0]  & [11$\bar{1}$] & [112] & -4.42 & -2.11 & -3.27 & -25.10  & +22.38  &-1.36 \\
		\hline\addlinespace[2pt]
		\bottomrule		
	\end{tabular}
\end{table*}

\subsection{Bending Response of Si NWs}
\label{sec:3.2}

\begin{figure}[ht]
	\centering
	\includegraphics[width=0.5\textwidth]{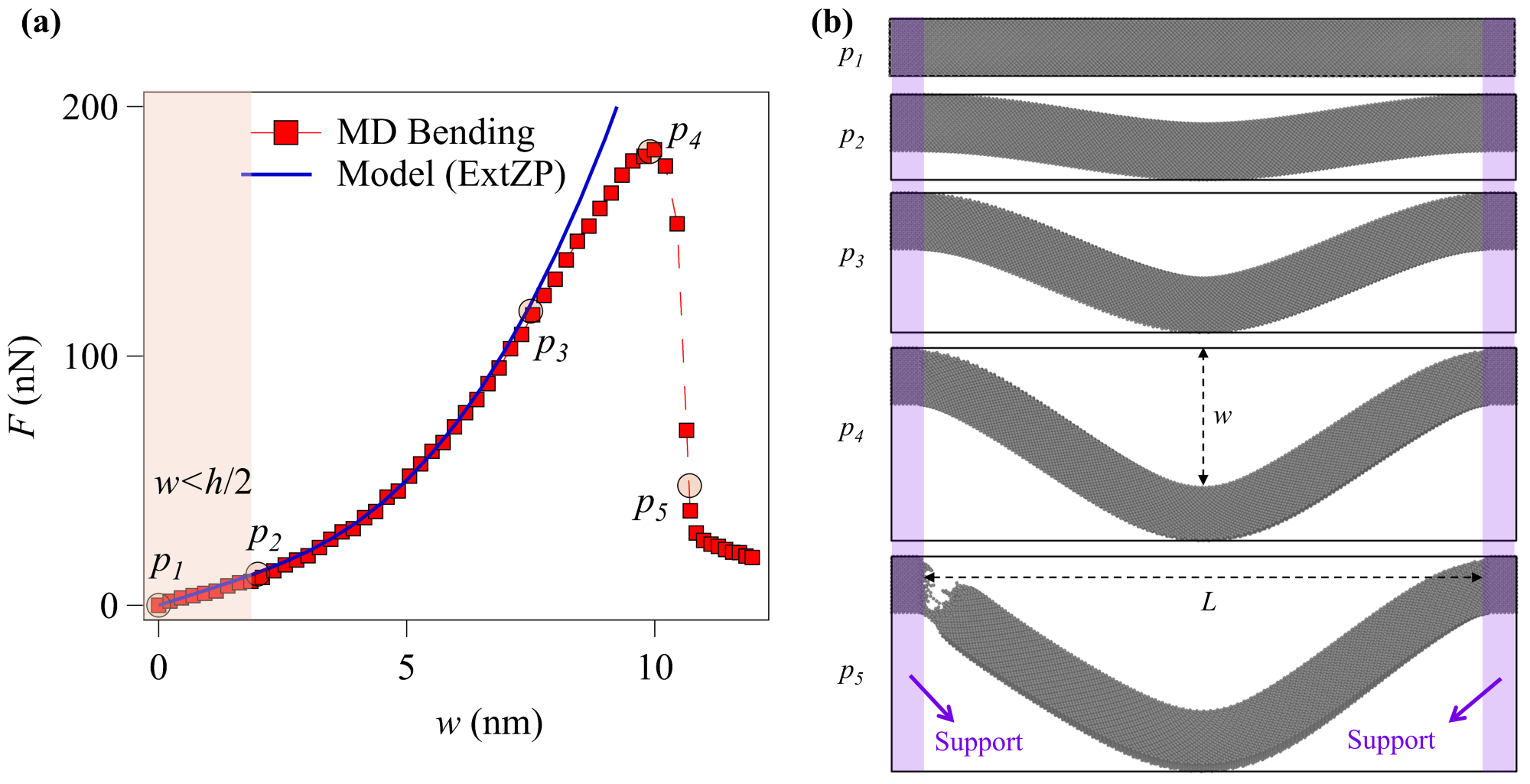}
	\caption{\label{fig:fig4} (a) Force vs. deflection data obtained through 3-point bending simulation of a Si NW of $h$ = 4 nm and fitted curve using the ExtZP model. (b) Atomic configurations of a Si NW and snapshots upon deformation of the Si NW during bending experiment.}
\end{figure}

In Figure~\ref{fig:fig4} (a), the graph demonstrates the correlation between force and deflection ($F-w$) during a three-point bending test of a Si NW simulated using the MD method. Five significant deformation points ($p_1$ to $p_5$) are identified for detailed examination of the $F-w$ behavior, and accompanying snapshots of the bending deformation of the Si NW profile are presented in Figure~\ref{fig:fig4} (b). The gray shaded region (between points $p_1$ and $p_2$) depicted in Figure~\ref{fig:fig4} (a) represents the range of deformation occurring in the middle of the Si NW, reaching approximately $h/2$, before transitioning into non-linearity, indicating large deflection during Si NW deformation. Likewise, Figure~\ref{fig:fig4} (b) illustrates the deflection level attained at $p_2$ as the bending continues in the mid-section, eventually reaching the ultimate force level. The point $p_3$ exhibits the deviation observed between nanomechanical modeling and $F-w$ response given here. The ultimate force is also shown as point $p_4$. Finally, the fracture snapshot corresponding to Si NW failure is denoted as $p_5$. The response of the Si NW demonstrates a nonlinear relationship between force and displacement, covering a force range of up to 180 nN and a corresponding deflection range of up to 10 nm at the center of the NW. Examination of Si NWs with different crystal orientations and critical dimensions shows that incorporating surface properties like surface stresses and surface elastic constants (given in Table~\ref{tab:table2}), along with intrinsic stresses, into the relevant models becomes essential to effectively apply various nanomechanical models for fitting the obtained $F-w$ responses. In this respect, MD simulations are utilized to determine the intrinsic stresses induced in Si NWs of different sizes and orientations. Once the relaxation process reaches a steady state, the atomic stresses along the longitudinal direction ($x$-axis) are computed for each Si NW using Equation~\ref{eqn18}. Table~\ref{tab:table3} presents the calculated surface-induced residual stresses for Si NWs with varying critical dimensions and orientations. In this study, it is assumed that the surface-induced residual stress represents the total intrinsic stress which will also be regarded as an initial tensile/compressive stress within the associated models. Intrinsic stresses in Si NWs have been reported in previous studies, including both computational \cite{zhan2012modified} and experimental \cite{zare2023nanomechanical, calahorra2015young, nasr2022effect, dolabella2020real, spejo2020non} investigations. These studies have observed intrinsic stress values in Si NWs within a range of 100 MPa to 1.2 GPa. The nanomechanical model fitting shown in Figure~\ref{fig:fig4} (a) corresponds to the ExtZP model, which is obtained by incorporating the relevant surface and intrinsic parameters. The detailed estimation of the modulus of elasticity for Si NWs with various dimensions and crystallographic orientations using nanomechanical models, as described in Section~\ref{sec:2.1}, is provided in Section~\ref{sec:3.3}.

\begin{table*}\centering
	\caption{\label{tab:table3} Intrinsic stresses of Si NWs with different critical dimensions and crystallographic orientations obtained via MD simulations.}
	\begin{tabular}{cccccccc}
		
		\toprule
		Structure & \multicolumn{3}{c}{Si Orientation} & \multicolumn{4}{c}{Intrinsic Stress [GPa]}  \\
		\midrule
		\hline\addlinespace[2pt]
		\hline\addlinespace[2pt]
		& $x$ & $y$ & $z$	& $h = 2 nm$ & $h = 3 nm$ & $h = 4 nm$	& $h = 5 nm$ \\
		\hline\addlinespace[2pt]
		\cmidrule(l){2-4}
		\cmidrule(l){5-8}
		Si $<100>$ 		 & {[100]}  & [010] & [001] & 		0.69 & 0.51 & 0.53 & 0.38   \\
		\hline\addlinespace[2pt]
		Si $<110>$      &	[110] & [001] &   [$\bar{1}$10] & 1.87 & 1.46 & 1.23 & 0.77  \\
		\hline\addlinespace[2pt]
		Si $<111>$ 		& [111] & [1$\bar{1}$0]  &  [$\bar{1}$$\bar{1}$2] &	1.73 & 0.98  &  1.26 & 0.80\\
		\hline\addlinespace[2pt]
		Si $<112>$  & [112]	  & [11$\bar{1}$] & [1$\bar{1}$0]  & 1.48 & 0.81 & 1.05 & 0.65  \\
		\hline\addlinespace[2pt]
		\bottomrule		
	\end{tabular}
\end{table*}

\subsection{Interpretation of Results}
\label{sec:3.3}

\begin{figure}[ht]
	\centering
	\includegraphics[width=0.5\textwidth]{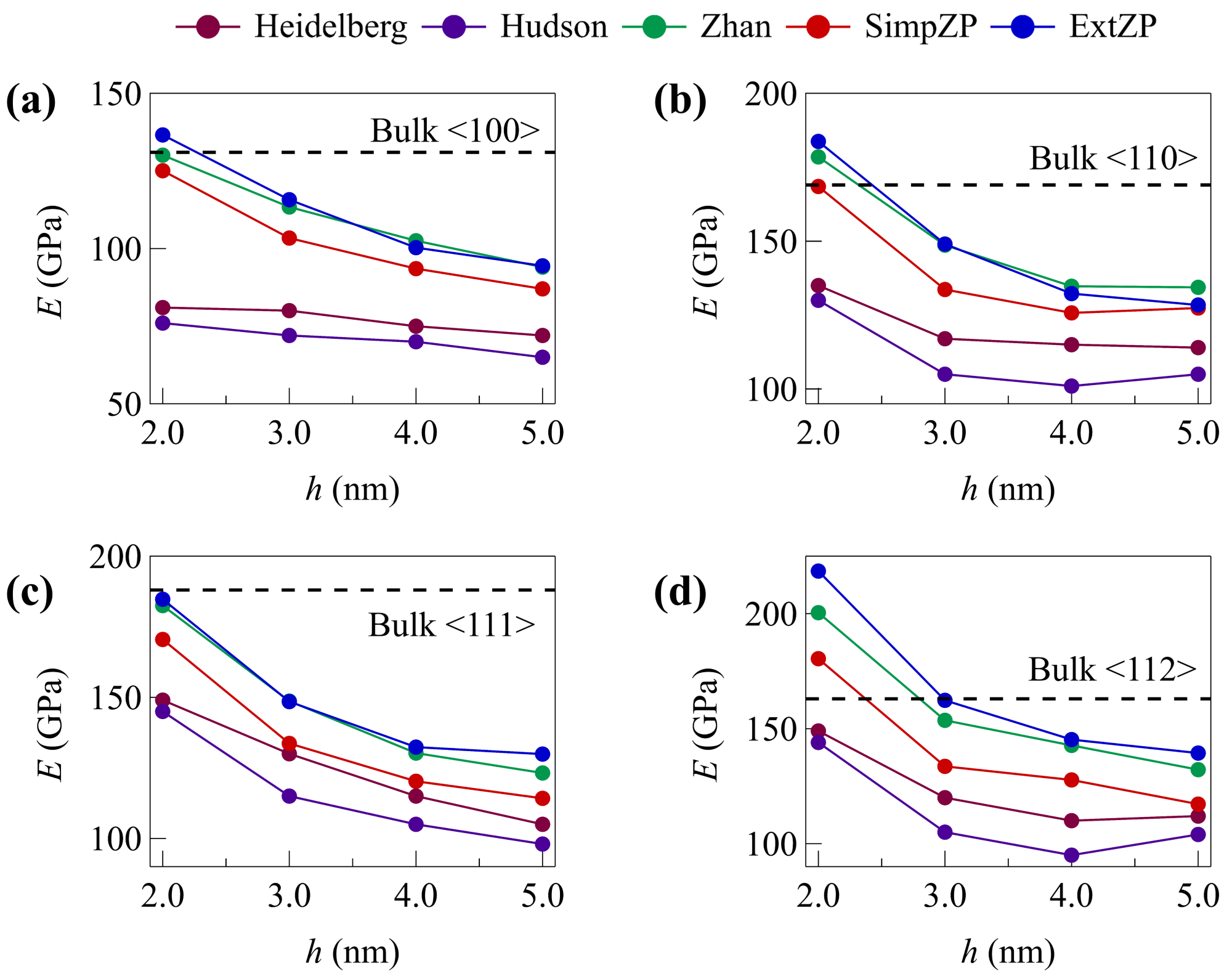}
	\caption{\label{fig:fig5} : The modulus of elasticity estimations for Si NWs with (a) $<100>$, (b) $<110>$, (c) $<111>$, and (d) $<112>$ crystal orientations. The Si NWs have different size in $h$ ranging between 2 to 5 nm. The bulk modulus of elasticity for each orientation of Si is shown.}
\end{figure}

This section provides a detailed explanation of the interpretation of force-deflection responses for Si NWs with different dimensions and crystal orientations. Previous studies \cite{heidelberg2006generalized, zhan2012modified, zare2023nanomechanical} have used a fitting procedure that begins with small deflections and extends to include progressively larger displacements. It is important to note that each nanomechanical model attributes the nonlinearity of the $F-w$ curve to different sets of parameters, including axial extension, intrinsic stress, and surface effects. Thus, each obtained $F-w$ curve is analyzed using five nanomechanical models, utilizing the appropriate input parameters, to obtain the modulus of elasticity of the Si NW. Among the models used, the ExtZP model is the comprehensive method that incorporates the surface stress ($f_{11}$ and $f_{22}$) and surface elastic constants ($d_{11}$ and $d_{22}$) provided in Table~\ref{tab:table2}, as well as the intrinsic stresses given in Table~\ref{tab:table3}, along with geometric details for interpreting the bending response. The SimpZP model considers the intrinsic stresses provided here as the initial tension ($\sigma_{0}$) and incorporates the surface properties $f_{11}$ and $d_{11}$ during the fitting procedure. The Zhan model relies on the previously given surface constants ($f_{11}$ and $d_{11}$) without including any intrinsic terms. In contrast, the Hudson model includes the intrinsic term based on the initial tension adopted from Table~\ref{tab:table3}, without incorporating any surface properties. Finally, the Heidelberg model performs the fitting procedure without including any surface or intrinsic terms. The bending rigidity formulations provided in Section~\ref{sec:2.2} are utilized with the previously mentioned details specific to each nanomechanical model, taking advantage of the relevant surface properties. The procedure outlined for estimating the modulus of elasticity using the provided models and the fitting curve shown in Figure~\ref{fig:fig4} (a) has been applied to Si NWs. The results are presented in Figure~\ref{fig:fig5} for different crystal orientations. Specifically, the modulus of elasticity estimation for $<100>$, $<110>$, $<111>$, and $<112>$-oriented Si NWs, with $h$ ranging from 2.0 to 5.0 nm, is shown in Figure~\ref{fig:fig5} (a), (b), (c), and (d), respectively. The bulk modulus of elasticity for each corresponding orientation is also given as dashed lines \cite{nasr2019review}. 

The trend observed for Si NWs with varying crystal orientations is a decrease in the modulus of elasticity as the critical dimension $h$ increases, though the rate of decrease varies depending on the orientation and the specific nanomechanical model employed. Specifically, for $<100>$-oriented Si NWs given in Figure~\ref{fig:fig5} (a), the modulus of elasticity estimation remains relatively constant for different critical dimensions, while incorporating the Heidelberg and Hudson models. When incorporating the surface and/or intrinsic terms in the Zhan, SimpZP, and ExtZP models, there is a consistent decreasing trend of $E$ as the critical dimension $h$ increases. The differences observed among these three models can be attributed to the utilization of distinct combinations of inputs, despite sharing a similar overall decreasing trend. For $<110>$-oriented Si NWs given in Figure~\ref{fig:fig5} (b), different decreasing trends are observable using the nanomechanical models. With the Heidelberg and Hudson model having higher estimation at 2.0 nm critical dimension, a similar constant trend is obtained with increase in $h$. However, the Zhan, SimpZP, and ExtZP models depict a decreasing $E$ for Si NWs with increasing $h$.

Expanding the discussion to $<111>$-Si NWs, as shown in Figure~\ref{fig:fig5} (c), a clear decreasing trend of the modulus of elasticity can be observed for all the nanomechanical models used in this study. The estimations consistently demonstrate a similar trend when considering the effect of critical dimensions on the Si NWs. However, similar to the previous orientations, the Zhan, SimpZP, and ExtZP models yield higher estimations of $E$ compared to the Heidelberg and Hudson models. For $<112>$-Si NWs, as shown in Figure~\ref{fig:fig5} (d), the trend of the modulus of elasticity also exhibits a similar decrease, but with differences between the models at each critical dimension. Once again, the Heidelberg and Hudson models provide lower estimations compared to the other three models. It is worth noting that the combination of parameters included in the nanomechanical models for the four crystal orientations studied here can result in differences of up to 100 GPa in the estimation of $E$, which is a significant variation. 
In particular, in the case of the smallest critical dimension ($h$ of 2.0 nm), where surface effects are anticipated to exhibit a more significant influence due to the size effect, models that encompass surface properties yield higher estimations, displaying more substantial deviations when contrasted with the Heidelberg and Hudson models. Notably, both the Heidelberg and Hudson models lack formulations that include surface-related terms, leading to inadequate estimations of the modulus of elasticity for Si NWs with critical dimensions below 10 nm. As the size increases, the surface contribution becomes less pronounced, leading to closer estimations among different models.
 
In the investigation of size-dependent elastic properties of Si NWs, especially in cases of smaller dimensions, it becomes crucial to meticulously account for the inclusion of terms, particularly those relating to surface properties. When comparing the outcomes of the three models incorporating surface terms, it becomes apparent that both the Zhan and SimpZP models employ a common surface model known as the modified YL model. Depending on whether intrinsic terms are taken into account or not, divergences emerge between these two models. The Zhan model consistently yields higher estimates of the modulus of elasticity for all four orientations and critical dimensions. Conversely, the SimpZP model, which incorporates a tensile intrinsic stress during the fitting procedure, is anticipated to offer lower estimates of $E$ depending on the critical dimension and the magnitude of the intrinsic stress. Furthermore, it is important to emphasize that among the four orientations studied in this work ($<110>$, $<111>$, $<112>$), only the ExtZP model benefits from the inclusion of anisotropic surface constants modeled via extended YL model given in Table~\ref{tab:table2}. The remaining four models dismiss some or all of these surface terms, depending on their respective formulations. Previous findings emphasizing the significance of the native oxide surface state \cite{zare2023nanomechanical, pakzad2021molecular, pakzad2023role, zare2023mechanical} can be extended via present study on unreconstructed Si NWs. This highlights not only the importance of the surface state itself but also the anisotropic effects of surface properties. The selection of an appropriate model is crucial to avoid potential misinterpretations, as improper model selections may lead to misleading conclusions regarding the nature of unreconstructed Si surfaces and their impact on Si NW behavior. Given the discrepancies observed in the literature regarding experimental and computational estimations of the modulus of elasticity in Si NWs, the values obtained in this study are in line with the prior reports \cite{zare2023nanomechanical, nasr2019review, wang2017mechanical, yang2022review}.

\section{Conclusion}
\label{sec:4}

This study exhibits the pivotal role that surface properties play in the mechanical behavior of Si NWs. By delving into the elastic properties of Si NWs with varying dimensions and crystal orientations through a diverse range of nanomechanical models, the intricate relationship between elastic properties, crystal orientation, and the chosen nanomechanical model has been unveiled. The analysis of force-deflection responses via MD simulations yields invaluable insights into the behavior of Si NWs at the nanoscale. These findings underscore the significance of meticulously modeling surface effects and incorporating anisotropic surface properties to attain a comprehensive grasp of Si NW mechanics. Additionally, the distinct trends observed in Si NWs with different crystal orientations emphasize the necessity of accounting for crystal orientation effects when scrutinizing their mechanical properties. This knowledge contributes to the ongoing exploration of size-dependent behaviors in Si NWs and enhances our comprehension of their surface attributes. A thorough comprehension of the effects originating from surface and crystal orientation aspects will undoubtedly expedite the development of more efficient and reliable NW-based technologies in the future.

\section*{ACKNOWLEDGMENT}

S.Z.P. and B.E.A. gratefully acknowledge the financial support by Tubitak under grant no. 120E347.

\bibliographystyle{IEEEtran}
\bibliography{LDM_SZP}

\end{document}